\documentstyle[12pt,epsf]{article}
\renewcommand{\bar}[1]{\overline{#1}}

\begin{document}

\begin{flushright}
GEF-Th-3/2002\\
\end{flushright}

\bigskip\bigskip
\begin{center}
{\large \bf The Transversity Function\\
and Double Spin Asymmetry in Semi-Inclusive DIS}
\end{center}
\vspace{12pt}

\begin{center}
 {\bf Elvio Di Salvo\\}

 {Dipartimento di Fisica and I.N.F.N. - Sez. Genova, Via Dodecaneso, 33 \\-
 16146 Genova, Italy\\}
\end{center}

\vspace{10pt}
\begin{center} {\large \bf Abstract}

We show that the spin asymmetry in semi-inclusive deep inelastic scattering 
between a longitudinally polarized lepton beam and a transversely polarized 
proton target is sensitive to the proton transversity function. The asymmetry, 
which is twist 3, is estimated to be more than $10\%$ under the most favourable 
conditions. The experiment we suggest is feasible at facilities like DESY, CERN 
and JLAB.

\end{center}

\vspace{10pt}

\centerline{PACS numbers: 13.85.Qk, 13.88.+e}

\newpage

$~~~~$ High energy spin physicists have been focusing for some time 
their efforts on determining the quark transversity distribution[1-5], which 
appears a particularly difficult task[6-10]. Indeed, different observables have 
been singled out, which are sensitive to this quantity; among them the 
Drell-Yan double spin asymmetry\cite{ba} and the interference fragmentation 
functions\cite{ja1}.    

For the moment the most promising experiments in this sense are those realized 
or planned by SMC\cite{smc} and HERMES\cite{her,he1} collaborations. These are 
based on the Collins effect\cite{coll} and consist in semi-inclusive 
pion electroproduction by a longitudinally\cite{her} or transversely\cite{he1}
polarized proton target. Up to now such experiments have provided a rough 
evaluation\cite{ef}
of the transversity function, $h_1^f$, where $f$ is the quark flavor. The 
single spin asymmetry is proportional to the product $h_1^f(x) c_f(z)$, where 
$c_f$ is the azimuthal asymmetry fragmentation function of a transversely 
polarized quark into a pion\cite{ja01,ef}, and, as usual, $x$ and $z$ are the 
longitudinal fractional momenta, respectively, of the active quark and of the 
pion with respect to the fragmenting quark. As claimed by Jaffe\cite{ja01}, 
this may become the "classic" way of determining the nucleon transversity 
distribution functions, provided $c_f(z)$ is known to some precision and not 
too small. But at present we know very little about this function\cite{ef}. 
Analogous considerations could be done about the method suggested by Jaffe and 
Ji (JJ)\cite{jj2}. This method consists in measuring the double spin asymmetry 
in a semi-inclusive deep inelastic scattering (SIDIS) experiment of the type 
\begin{equation}
\vec{\ell} p^{\uparrow} \to \ell' \pi X,
\label{r1}
\end{equation}
where $\vec{\ell}$ is a longitudinally polarized charged lepton and 
$p^{\uparrow}$ a transversely polarized proton target. The asymmetry is defined 
as
\begin{equation}
A(|{\bf k}|; Q, \nu; \Pi_{\parallel}) = \frac{d\sigma_{\uparrow \rightarrow} - 
d\sigma_{\uparrow \leftarrow}}{d\sigma_{\uparrow \rightarrow} + 
d\sigma_{\downarrow \leftarrow}}.
\label{as1}
\end{equation}
Here, as usual, $\nu$ is the lepton energy transfer and $Q^2$ = $-q^2$, $q$ 
being the four-momentum transfer. Furthermore ${\bf k}$ is the momentum of the 
initial lepton and $\Pi_{\parallel}$ the component of the final pion momentum 
along the momentum transfer. Lastly
$d\sigma_{\uparrow\rightarrow}$ and $d\sigma_{\uparrow\leftarrow}$ are the
polarized differential cross sections for reaction (\ref{r1}), integrated over 
the transverse momentum of the final pion with respect to the momentum 
transfer; arrows indicate the proton and lepton polarization. Asymmetry 
(\ref{as1}) contains a twist-4 term\cite{jj2} which includes the product 
$h_1^f(x){\hat e}^f(z)$, where ${\hat e}^f(z)$ is the twist-3 fragmentation 
function of the pion. Again, the extraction of $h_1^f$  depends critically on 
an unknown function.

Alternatively, in this letter we consider the double spin asymmetry $A(|{\bf 
k}|; Q, \nu; {\bf P})$, analogous to (\ref{as1}), but where the pion momentum
${\bf P}$ is kept fixed. As we shall see, such an asymmetry - which is of the 
type considered by Mulders and Tangerman\cite{mt} (MT) - is twist 3 and 
sensitive to the transverse momentum dependent (t.m.d.) transversity function,
without involving any unknown distribution or fragmentation functions. The 
t.m.d. transversity function depends on $x$, on ${\bf p}_{\perp}^2$ and on
$({\bf p}_{\perp}\cdot{\bf S})^2$, where ${\bf p}_{\perp}$ is the quark 
transverse 
momentum and ${\bf S}$ the proton spin vector, ${\bf S}^2$ = 1. In a short-hand 
notation, we indicate this function as $\delta q_{\perp}(x, {\bf p}_{\perp})$.

As is well-known\cite{coll}, SIDIS is 
kinematically isomorphic to Drell-Yan. But we have shown in a previous 
paper\cite{dis} that the inclusive muon pair production from singly polarized 
proton-hadron collisions, at a fixed transverse momentum of the pair with 
respect to the initial beams, causes a muon polarization, which is sensitive 
to $\delta q_{\perp}$. Therefore we expect an analogous effect in reaction 
(\ref{r1}), provided we fix the pion direction. This is indeed the case,
as we shall see.

We calculate $A(|{\bf k}|; Q, \nu; {\bf P})$ in the framework of a 
QCD-improved parton model\cite{si}. In the laboratory frame the differential 
cross section reads, in one-photon exchange approximation,
\begin{equation}
d\sigma = {1 \over {4 |{\bf k}| m}} {{e^4} \over {Q^4}} L_{\mu\nu} H^{\mu\nu} 
\ d\Gamma,
\label{dsg}
\end{equation}
where $m$ is the proton rest mass and $L_{\mu\nu}$ ($H_{\mu\nu}$) the leptonic 
(hadronic) tensor. $d\Gamma$, the phase space element, reads
\begin{equation}
d\Gamma = \frac{1}{(2\pi)^6} d^4 k ~ \delta(k^2) ~ \theta(k_0) ~~ 
d^4 P ~ \delta(P^{2}-m_{\pi}^2) ~ \theta(P_0). 
\label{mps}
\end{equation}
Here $k$ and $P$ are, respectively, the four-momenta of the initial lepton and 
of the pion, whose rest mass is $m_{\pi}$. 
The leptonic tensor is, in the massless approximation, 
\begin{equation}
L_{\mu\nu} = \frac{1}{4} Tr[\rlap/k(1+\lambda_{\ell}\gamma_5)
\gamma_{\mu}\rlap/k'\gamma_{\nu}],
\label{lept}
\end{equation}
$\lambda_{\ell}$ being the helicity of the initial lepton and $k'$ 
= $k-q$ the four-momentum of the final lepton. Trace calculation yields
\begin{equation}
L_{\mu\nu} = k_{\mu} k'_{\nu} + k'_{\mu} k_{\nu} - g_{\mu\nu} k \cdot k' + 
i \lambda_{\ell}\varepsilon_{\alpha\mu\beta\nu} k^{\alpha} k^{'\beta}.
\label{lept1}
\end{equation}
As regards the hadronic tensor, 
the generalized factorization theorem\cite{qi1,tm,bo} in the covariant 
formalism\cite{land} yields, at zero order in the QCD coupling constant, 
\begin{eqnarray}
H_{\mu\nu} &=& \frac{1}{3} \sum_{f=1}^6 e_f^2 \int d\Gamma_q \varphi^{f} (p';P) 
h^f_{\mu\nu} (p, p'; S), \label{hadt} 
\\
h^f_{\mu\nu} &=& \sum_Lq^f_L(p) Tr (\rho^L \gamma_{\mu}\rho' \gamma_{\nu}).
\label{form0} 
\end{eqnarray}
Here the factor 1/3 comes from color averaging in the elementary scattering 
process and $f$ runs over the three light flavors ($u,d,s$) and antiflavors 
($\bar{u},\bar{d},\bar{s}$),  $e_1$ = $-e_4$ = 2/3, $e_2$ = $e_3$ = $-e_5$ = 
$-e_6$ = -1/3. $p$ and $p'$ are respectively the four-momenta of 
the active parton before and after being struck by the virtual photon. $S$ is 
the Pauli-Lubanski (PL) four-vector of the proton. $q^f_L$ is the probability 
density function of finding a quark 
(or an antiquark) in a pure spin state, whose third component along the proton
polarization is $L$. Analogously $\varphi^f$ is the fragmentation function of a 
quark of four-momentum $p'$ into a pion of four-momentum $P$. Moreover
\begin{equation}
d\Gamma_q = \frac{1}{(2\pi)^2} d^4 p ~ \delta(p^2) \theta(p_{0}) 
d^4 p' ~ \delta(p^{'2}) \theta(p'_{0}) ~~ \delta^4(p'-p-q),
\label{qps}
\end{equation}
the active parton being taken on shell and massless. Lastly the $\rho$'s are 
the spin density matrices of the initial and final active parton, 
{\it i. e.}\cite{dis},
\begin{equation}
\rho^L = {1 \over 2} \rlap/p [1 + 2 L\gamma_5 (\lambda + 
\rlap/\eta)] \ ~~~~~~ {\mathrm and} ~~~~~~~ \
\rho' = {1 \over 2} \rlap/p'. 
\label{dens}
\end{equation}
Here $2L\eta$ is the transverse PL four-vector of the active parton, while 
$\lambda$ is the longitudinal component of the quark spin vector. Formulae 
(\ref{dens}) are consistent with the Politzer theorem\cite{po} in the parton 
model approximation. These imply, together with eq. (\ref{form0}), 
that $\eta$ does not contribute to $h^f_{\mu\nu}$.
For later convenience we re-write this last tensor as 
\begin{equation}
h^f_{\mu\nu} = \frac{1}{4}\left[q^f(p) s_{\mu\nu}+\lambda\delta q^f(p) 
a_{\mu\nu}\right], \label{form1}
\end{equation}
where
\begin{equation}
s_{\mu\nu} = Tr(\rlap/p \gamma_{\mu}\rlap/p'\gamma_{\nu}), \ ~~~~~ \ ~~~~~~ \ 
a_{\mu\nu} = Tr(\gamma_5\rlap/p \gamma_{\mu}\rlap/p'\gamma_{\nu})
\label{form13}
\end{equation}
and $q^f(p) = \sum_Lq^f_L(p)$ is the unpolarized quark distribution function, 
while $\delta q^f(p) = \sum_L2Lq^f_L(p)$. 

$\lambda$ is a Lorentz scalar, such that $|\lambda|$ $\leq$ 1. 
If we neglect the parton transverse momentum,
the only way of constructing such a quantity with the available vectors is
\begin{equation}
\lambda = \lambda_{\parallel}  = {\bf S} \cdot \frac{\bf q}{|{\bf q}|} = 
\frac{-S \cdot q}{\sqrt{\nu^2+Q^2}}. \label{elic}
\end{equation}
Here we have exploited the fact that $\nu$ is a Lorentz scalar and that in the 
laboratory frame $S$ $\equiv$ $(0,{\bf S})$ and $q$ $\equiv$ $(\nu,{\bf q})$. 
$\lambda_{\parallel}$ can be viewed as the helicity of the proton in a 
frame moving along ${\bf q}$. Now, in order to take into account the
transverse momentum, we have to adopt a frame where the proton momentum is 
large in comparison to $m$\cite{ale}. But, in this more refined approximation, 
we are still faced with the problem of defining $\lambda$ in a Lorentz 
invariant way. As we are going to show, the only way to do this is to consider 
the Breit frame, that is, where the virtual photon has four-momentum $q$ = 
$(0, {\bf q}_B)$, with $|{\bf q}_B|$ = $Q$. In this frame the proton momentum 
is $-\frac{1}{2x} {\bf q}_B$, therefore the active parton carries a momentum 
${\bf p}_B = -\frac{1}{2} {\bf q}_B + {\bf p}_{\perp}$, where, as usual, $x = 
Q^2/(2m\nu)$ is the longitudinal fractional momentum and ${\bf p}_{\perp}$ the 
transverse momentum with respect to ${\bf q}_B$. 
We decompose the proton spin vector ${\bf S}$ into a longitudinal and a 
transverse component, {\it i. e.},
\begin{equation}
{\bf S} = \lambda_{\parallel}\frac{\bf q}{|{\bf q}|} + {\bf S}_{\perp}, \ 
~~~~~~ \ ~~~~~ \ {\bf S}_{\perp} \cdot {\bf q} = 0. \label{rel111}
\end{equation}
Moreover we carry on the integration (\ref{hadt}) over the time and 
longitudinal components of $p$, taking the $z$-axis opposite to ${\bf q}$. We 
get, in the light cone formalism,
\begin{equation}
H_{\mu\nu} = \frac{1}{4\pi^2 Q^2} \sum_{f=1}^6 e_f^2 \int d^2 p_{\perp} 
\varphi^{f} (z, {\bf P}^{2}_{\perp}) h^f_{\mu\nu} (x, {\bf p}_{\perp}; 
{\bf S}), 
\label{hadt2}
\end{equation}
where, taking into account the decomposition (\ref{rel111}) of ${\bf S}$, 
the tensor $h^f_{\mu\nu}$ (see eq. (\ref{form1})) reads\cite{jj}
\begin{equation}
h^f_{\mu\nu} = \frac{1}{4}\left\{q^f(x, {\bf p}_{\perp}^2) s_{\mu\nu}+
\left[\lambda_{\parallel}\delta q_{\parallel}^f(x, {\bf p}_{\perp}^2)+
\lambda_{\perp}\delta q_{\perp}^f(x, {\bf p}_{\perp})\right] a_{\mu\nu}\right\}.
\label{elemt}
\end{equation}
Here 
\begin{equation}
\lambda_{\perp} = \frac{{\bf S}_{\perp}\cdot {\bf p}_B}{|{\bf p}_B|} \simeq
\frac{2{\bf S}\cdot {\bf p}_{\perp}}{Q} \label{helic33}
\end{equation}
and $\delta q_{\parallel}^f(x, {\bf p}_{\perp}^2)$ is the t.m.d. helicity 
distribution function. Moreover $z$ = $(P_{\parallel}+P_0)/(2|{\bf p}'|)$ is 
the longitudinal fractional momentum of the pion resulting from fragmentation 
of the struck parton, whose momentum is ${\bf p}'$. We have defined $P_0$ = 
$\sqrt{m_{\pi}^2+{\bf P}^2}$, $P_{\parallel}$ = ${\bf P}\cdot{\bf p}'/|{\bf 
p}'|$ and ${\bf P}_{\perp}$ = ${\bf P}-P_{\parallel}{\bf p}'/|{\bf p}'|$. 
Denoting by ${\bf \Pi}_{\perp}$ the transverse 
momentum of the pion with respect to the photon momentum, we get
\begin{equation}
{\bf P}_{\perp} = {\bf \Pi}_{\perp}-z{\bf p}_{\perp}.
\label{rel002}
\end{equation}
Therefore, if we 
keep ${\bf \Pi}_{\perp}$ fixed, ${\bf P}_{\perp}$ depends on ${\bf p}_{\perp}$.
Since we want to pick up a pion resulting from fragmentation of the active 
quark, we pick up events such that $|{\bf \Pi}_{\perp}|$ $<<$ $|{\bf P}|$. 
Notice that, 
although we have chosen a particular frame - coincident with the one adopted by 
Feynman\cite{fey} -, the tensor (\ref{elemt}) is covariant.

In order to calculate the asymmetry $A(|{\bf k}|; Q, \nu; {\bf P})$ - defined 
analogously to (\ref{as1}), but keeping ${\bf P}$ fixed - we have to substitute 
the leptonic tensor (\ref{lept1}) and the hadronic tensor (\ref{hadt2}) into 
the cross section (\ref{dsg}), taking into account relations (\ref{elemt}) and 
(\ref{helic33}). The result is 
\begin{equation}
A(|{\bf k}|; Q,\nu; {\bf P}) = {\cal F} ~~ 
\frac{\sum_{f=1}^6 e_f^2\delta Q^f}
{\sum_{f=1}^6 e_f^2 Q^f}, 
~~~~ \ ~~~~
{\cal F} = \frac{k_+k'_- - k_-k'_+}{k_+k'_- + k_-k'_+}. \label{assidis} 
\end{equation} 
Here we have introduced the quantities
\begin{eqnarray}
Q^f &=& Q^f(x,z,{\bf \Pi}_{\perp}^2) = \int d^2 p_{\perp} q^f(x, 
{\bf p}_{\perp}^2) \varphi^f (z, {\bf P}^{2}_{\perp}), 
\label{qf}
\\
\delta Q^f &=& \delta Q_{\parallel}^f(x,z,{\bf \Pi}_{\perp}^2) +  
{\bf \Pi}_{\perp}\cdot {\bf S} 
\delta Q_{\perp}^f(x,z,{\bf \Pi}_{\perp}), \label{dqf}
\end{eqnarray}
\begin{eqnarray}
\delta Q_{\parallel}^f(x,z,{\bf \Pi}_{\perp}^2) &=& \lambda_{\parallel}
\int d^2 p_{\perp} \delta q^f_{\parallel} (x, {\bf p}_{\perp}^2) \varphi^f (z, 
{\bf P}^{2}_{\perp}), 
\label{dqf0}
\\
\delta Q_{\perp}^f(x,z,{\bf \Pi}_{\perp}) {{\bf \Pi}_{\perp}\cdot {\bf S}} &=& 
\int d^2 p_{\perp} \lambda_{\perp}
\delta q^f_{\perp} (x, {\bf p}_{\perp}) \varphi^f (z, {\bf P}^{2}_{\perp}).
\label{dqf1}
\end{eqnarray}
Below we shall explain the presence of the scalar product ${{\bf 
\Pi}_{\perp}\cdot {\bf S}}$ at the left side of eq. (\ref{dqf1}) and we shall 
show that $\delta Q^f$ is twist 3. Since the products $k_+ k'_-$ and $k_- k'_+$ 
are invariant under boosts along the $z$-axis, we calculate them in the 
laboratory frame, where, in the massless approximation, 
\begin{equation}
k_{\pm} = \frac{|{\bf k}|}{\sqrt{2}}(1\pm cos\beta), ~~~~~~ \ ~~~~~ \ 
k'_{\pm} = \frac{|{\bf k'}|}{\sqrt{2}}[1\pm cos(\theta+\beta)].
\label{pmc}
\end{equation} 
Here ${\bf k}' = {\bf k}-{\bf q}$ is the final lepton momentum. Moreover
$\beta$ and $\theta$ are, respectively, the angle between ${\bf k}$ and {\bf q}
and between ${\bf k}$ and ${\bf k}'$:  
\begin{equation}
|{\bf q}| cos\beta = |{\bf k}| - |{\bf k}'| cos\theta. \label{angg}
\end{equation} 
Now we consider the scaling limit, {\it i. e.},
$Q^2 \to \infty$, $\nu \to \infty$, $\frac{Q^2}{2m\nu} \to x$.
Since $Q^2 \simeq 2 |{\bf k}| |{\bf k}'| (1-cos\theta)$,
$\theta$ tends to zero in that limit, as well as $\beta$: 
\begin{equation}
\theta \simeq \frac{m}{Q} \frac{y}{x(1-y)^{1/2}}, ~~~~ \ ~~~~
\beta \simeq \theta\frac{1-y}{y}, ~~~~ \ ~~~~ y = \frac{\nu}{|{\bf k}|}. 
\label{rell2}
\end{equation} 
Then the second eq. (\ref{assidis}) and eq. (\ref{elic}) yield, respectively,
\begin{equation}
{\cal F} = \frac{y(2-y)}{1+(1-y)^2}, ~~~~ \ ~~~~ \lambda_{\parallel} = 
\frac{1-y}{y}sin\theta cos\phi, \label{kfac}
\end{equation} 
where $\phi$ is the azimuthal angle between the (${\bf k},{\bf k}'$) plane and 
the (${\bf k},{\bf S}$) plane. Therefore $\delta Q_{\parallel}^f$, eq. 
(\ref{dqf0}), is 
twist 3, as follows from the second eq. (\ref{kfac}) and from the first eq. 
(\ref{rell2}). But also the second term of eq. (\ref{dqf}) is twist 3, as 
is immediate to check. Therefore our asymmetry is twist 3. Some remarks are in 
order.

(i) Invariance of strong interactions under parity and time reversal, together
with a rotation by $\pi$ around the proton momentum, implies
\begin{equation}
\delta q^f_{\perp} (x, {\bf p}_{\perp}) = \delta q^f_{\perp}  
(x, -{\bf p}_{\perp}).
\label{symde}
\end{equation}
Therefore the second term of eq. (\ref{dqf}) vanishes either if we set ${\bf 
\Pi}_{\perp}$ = 0 or if we integrate over ${\bf \Pi}_{\perp}$. The former 
property explains the presence of the factor ${{\bf \Pi}_{\perp}\cdot {\bf S}}$ 
at the l.h.s. of eq. (\ref{dqf1}). From this we deduce that, in order to 
maximize our asymmetry, it is 
most convenient to take the vector ${\bf \Pi}_{\perp}$ parallel to or opposite
to ${\bf S}$, that is, to select pions whose momenta lie in the (${\bf q}$, 
${\bf S}$) plane. On the other hand, integrating  over ${\bf \Pi}_{\perp}$, 
$\delta Q^f$ goes over into $\lambda_{\parallel}\Delta q^f(x)D^f(z)$,
corresponding to the twist-3 term of the numerator in the JJ 
asymmetry\cite{jj2}. Here $\Delta q^f(x)$ is the helicity distribution function 
and $D^f(z)$ the usual fragmentation function of the pion. The above mentioned
numerator includes also a twist-4 term of the type 
$\lambda [\frac{1}{x}h_1^f(x)\frac{1}{z}{\hat e}^f(z)+g^f_T(x)D^f(z)]
$\cite{jj2}, where $g^f_T(x)$ is the transverse spin distribution function.

(ii) The second term of eq. (\ref{dqf}) is especially sensitive to 
$\delta q^f_{\perp} (x, {\bf p}_{\perp})$ if
${\bf q}\cdot{\bf S}$ = 0. In this situation the first term of eq. 
(\ref{dqf}) - and more generally the JJ asymmetry - vanishes. 
Therefore events such that the lepton scattering plane is perpendicular to the 
proton polarization, that is $\phi\ = \pi/2$, are particularly relevant to our 
aims.

(iii) It is worth observing that the t.m.d. transversity function has a 
chiral-even component, owing to the non-collinearity of the quark with respect
to the proton momentum. To see this, it is sufficient to change the 
quantization axis from the proton momentum  to the quark momentum. It is just 
this chiral-even component that appears in formula (\ref{dqf1}); indeed, as we 
have seen before, the quark density matrix involves only the helicity.
Therefore, according to chirality conservation, our asymmetry formula 
(\ref{assidis}) - unlike those previously considered 
for determining the transversity function[6-12] - 
does not contain any chiral-odd (and therefore unusual) distribution or 
fragmentation functions. The t.m.d. functions $q^f$ and $\varphi^f$, involved 
in the asymmetry, can be parametrized in a well defined way. Incidentally, the 
change in the quantization axis allows to establish connections between the
transverse momentum dependent distribution functions defined by MT (see also 
refs.\cite{tm} and\cite{rs}); for example, in the MT notation, $h_{1T} +
h_{1T}^{\perp} ({\bf S}_\perp\cdot {\bf p}_{\perp})^2/(m^2|{\bf S}_\perp|)$
- to be identified with $\delta q^f_{\perp} (x, {\bf p}_{\perp})$ -
is related to $g_{1T}$\cite{dis2}.

(iv) The asymmetry derives contributions also from the one-gluon 
exchange\cite{qi1} terms, demanded by gauge invariance.
In our approach these terms are conveniently calculated in the light 
cone gauge. Two of these terms, which may be also deduced from the equations of 
motion\cite{bo1}, result in the distribution $\delta g^f_T(x,{\bf 
p}_{\perp})$\cite{bo1,mt}, whose integral over the transverse momentum is 
$g^f_{T}(x)$. Calculations\cite{dis} within the model proposed by Qiu and 
Sterman\cite{qi1} assure that such contributions are about $10\%$ of the zero 
order term.

Lastly we calculate the order of magnitude of the asymmetry (\ref{assidis}) 
under optimal conditions. To this end, according to the above considerations, 
we take $\phi$ = $\pi/2$ and ${\bf \Pi}_{\perp}$ parallel or antiparallel to 
${\bf S}$. Moreover we assume\cite{mt,tm,bo} a gaussian 
behavior for the transverse momentum dependence of the functions involved in 
our asymmetry, {\it i. e.}, $q^f(x,{\bf p}_{\perp}^2)$ = $(\pi a)^{-1}q^f(x) 
exp(-a {\bf p}_{\perp}^2)$, and analogously for $\delta q^f_{\perp}(x,{\bf 
p}_{\perp})$ and $\varphi^f(z,{\bf P}_{\perp}^2)$, the parameter $a$ being 
taken equal for all three functions. The asymmetry (\ref{assidis}) results
in
\begin{equation}
A(y; Q, x; z, {\bf \Pi}_{\perp}) = \frac{{\bf S}\cdot{\bf \Pi}_{\perp}}{Q} 
\frac{2z{\cal F}}{1+z^2}\frac{\sum_{f=1}^6 e^2_f h_1^f(x)D^f(z)}
{\sum_{f=1}^6 e^2_f q^f(x)D^f(z)}.\label{fin}
\end{equation}
Eq. (\ref{fin}) and the first eq. (\ref{kfac}) suggest that the asymmetry is 
largest for $y$ and $z$ as close as possible to 1. Under such conditions, and
setting $|{\bf \Pi}_{\perp}|\simeq 1$ $GeV$ and $Q$ = 2.5 $GeV$, we have $A 
\sim 0.4 R$, where $R = h_1^f(x)/q^f(x)$ has been determined by 
HERMES\cite{her}, $|R| = (50\pm 30)\%$. 

To conclude, we have suggested a SIDIS experiment, using a longitudinally 
polarized lepton and a transversely polarized proton, and detecting 
a pion in the final state, at a not too large angle with ${\bf q}$. 
This experiment yields a double spin asymmetry sensitive to $h_1$ through 
$\delta q^f_{\perp}$, while the other functions involved in the asymmetry 
can be parametrized in a well defined way. For reasonable values of 
$Q^2$ (4 to 10 $GeV^2$), and under the most favourable kinematic conditions
(lepton scattering planes nearly orthogonal to ${\bf S}$ and pions whose 
momenta lie near the (${\bf q}$, ${\bf S}$) plane),
the order of magnitude of the asymmetry is estimated to be at least $\sim 10\%$.
The suggested experiment could be performed at 
facilities like CERN (COMPASS coll.), DESY (HERMES coll.) or Jefferson 
Lab., where similar asymmetry measurements are being realized or planned.

\end{document}